\documentclass{article}
\usepackage{wrapfig}
\usepackage{color}
\usepackage{graphicx}
\usepackage{amsmath,amsfonts,amsthm}

\makeatletter
\@addtoreset{equation}{section}
\makeatother

\def\supp{\mathrm{supp\,}}
\newtheorem{theorem}{Theorem}[section]

\newtheorem{exercise}{Exercise}[section]
\newtheorem{examps}{Examples}[section]
\newtheorem{lemma}{Lemma}[section]
\newtheorem{remark}{Remark}[section]

\newtheorem{proposition}{Proposition}[section]
\newtheorem{corollary}{Corollary}[section]
\newtheorem{definition}{Definition}[section]
\def\le{\left}
\def\ri{\right}

\def\ds{\displaystyle}

\def\res{\mathop{\mathrm {res}}\limits_}

\def\br{\begin{remark}}
\def\er{\end{remark}}
\def\bt{\begin{theorem}}
\def\et{\end{theorem}}
\def\bc{\begin{corollary}}
\def\ec{\end{corollary}}
\def\bx{\begin{examp}\small}
\def\ex{\end{examp}}
\def\bxr{\begin{exercise}\small}
\def\exr{\end{exercise}}
\def\bl{\begin{lemma}}
\def\el{\end{lemma}}
\def\bxs{\begin{examps}. \rm\begin{enumerate}}
\def\exs{\end{enumerate}\end{examps}}
\def\D{\mathbb D}
\def\bd{\begin{definition}}
\def\ed{\end{definition}}
\def\bp{\begin{proposition}}
\def\ep{\end{proposition}}
\def\be{\begin{equation}}
\def\ee{\end{equation}}

\def\&{\hspace{-15pt}&}
\def\bea{\begin{eqnarray}}
\def\eea{\end{eqnarray}}
\def\beas{\begin{eqnarray*}}
\def\eeas{\end{eqnarray*}}

\def \pa{\partial}
\def\C{{\mathbb C}}

\def\R{{\mathbb R}}
\def\N{{\mathbb N}}

\def\wh{\widehat}

\def\a{\alpha}
\def\d{\,\mathrm d}

\def\1{{\bf 1}}

\def\wt{\widetilde}
\def\ds{\displaystyle}

\def\tr{\mathrm {Tr}}

\date{}
\begin{document}
\baselineskip 16pt plus 1pt minus 1pt

\vspace{0.2cm}
\begin{center}
\begin{Large}
\textbf{Mesoscopic colonization of a spectral band}
\end{Large}\\
\bigskip
\begin{large} {M.
Bertola$^{\dagger,\sharp}$, S. Y. Lee$^{\ddagger}$, M. Y.
Mo$^{\flat}$}\footnote{Work supported in part by the Natural
    Sciences and Engineering Research Council of Canada
(NSERC). M. Y. Mo would like to acknowledge EPSRC for financial
support. (grant no.
EP/D505534/1)}\footnote{bertola@crm.umontreal.ca}
\end{large}
\\
\bigskip
\begin{small}
$^{\ddagger}$ {\em Department of Mathematics and
Statistics, Concordia University\\ 1455 de Maisonneuve W., Montr\'eal, Qu\'ebec,
Canada H3G 1M8} \\
$^{\sharp}$ {\em Centre de recherches math\'ematiques, Universit\'e\ de
Montr\'eal } \\
$^{\flat}$ {\em Department of Mathematics, University of Bristol,\\
University Walk
Bristol, BS8 1TW}\\
\end{small}
\bigskip
{\bf Abstract}
\end{center}
We consider the unitary matrix model in the limit where the size of the matrices become infinite and in the critical situation when a new spectral band is about to emerge. In previous works the number of expected eigenvalues in a neighborhood of the band was fixed and finite, a situation that was termed  ``birth of a cut'' or ``first colonization''.
We now consider the transitional regime where this microscopic population in the new band grows without bounds but at a slower rate than the size of the matrix. The local population in the new band organizes in a ``mesoscopic" regime, in between the macroscopic behavior of the full system and the previously studied microscopic one. The mesoscopic colony may form a finite number of new bands, with a maximum number dictated by the degree of criticality of the original potential. 
We describe the delicate scaling limit that realizes/controls the mesoscopic colony. 
The method we use is the steepest descent analysis of the Riemann-Hilbert problem that is satisfied by the associated orthogonal polynomials. 


\tableofcontents

\section{Introduction and result}
The phenomenon that we want to investigate in this paper goes under the name of ``birth of a cut'' \cite{EynardBirth,MoBirth, Claeys} or ``colonization of an outpost'' \cite{BertoLee1,BertoLee2}, namely the transition when one or more new spectral bands open in the asymptotic spectrum of the model. 
In particular we want to focus on the transition between the {\em microscopic} regime  (of finite number of eigenvalues) and the macroscopic regime (where the number of eigenvalues scales like $N$); we call the intermediate regime the {\em mesoscopic} regime. 

While the paper does not aim at being propaedeutic to the topic of random matrices, in this section we recall some general facts about the unitary random matrix model so as to set the context.  Unitary random matrix model is defined by the probability distribution
\begin{equation}\label{eq:rm}
Z_{n,N}^{-1}\exp\left(-\frac{N}{T}\tr V(M)\right)dM,\quad
Z_{n,N}=\int_{\mathcal{H}_n}\exp\le(-\frac{N}{T}\tr V(M)\ri)dM\ ,
\end{equation}
on the space $\mathcal{H}_n$ of Hermitian $n\times n$
matrices $M$, with $V$ a real analytic function (the {\bf potential}) that satisfies
\begin{equation*}
\lim_{x\rightarrow\pm\infty}\frac{V(x)}{\log(x^2+1)}=+\infty.
\end{equation*}
The eigenvalues $x_1,\ldots,x_n$ of the matrices in this ensemble are
distributed according to the probability distribution (See, e.g.
\cite{MehtaBook}, \cite{Dyson})
\begin{equation}\label{eq:proeig}
\mathcal{P}^{(n,N)}(x_1,\ldots,x_n)d^nx=\hat{Z}_{n,N}^{-1}e^{-\frac{N}{T}\sum_{j=1}^nV(x_i)}\prod_{j<k}(x_j-x_k)^2dx_1\ldots
dx_n,
\end{equation}
where $\hat{Z}_{n,N}$ is the normalization constant.

The correlation functions of the eigenvalues are related to orthogonal polynomials (see e.g. \cite{Dyson}, \cite{MehtaBook}): let $\{\pi_n(x)\}_{n\in\N}$ be the degree $n$
monic orthogonal polynomials with weight $e^{-NV(x)}$ on
$\mathbb{R}$. \cite{Szego}
\begin{equation}\label{eq:op}
\int_{\mathbb{R}}\pi_n(x)\pi_m(x)e^{-\frac{N}{T}V(x)}dx=h_n\delta_{nm}.
\end{equation}
Let us construct the correlation kernel by
\begin{equation*}
K_{n,N}(x,x^{\prime})=e^{-\frac{1}{2}\frac{N}{T}(V(x)+V(x^{\prime}))}\sum_{j=0}^{n-1}\frac{\pi_j(x)\pi_j(x^{\prime})}{h_j}.
\end{equation*}
By the Christoffel-Darboux formula, this kernel can be expressed in
terms of the two orthogonal polynomials $\pi_n(x)$ and
$\pi_{n-1}(x)$ instead of the whole sum:
\begin{equation}\label{eq:kernel}
K_{n,N}(x,x^{\prime})=e^{-\frac{1}{2}\frac{N}{T}(V(x)+V(x^{\prime}))}\frac{\pi_n(x)\pi_{n-1}(x^{\prime})-\pi_n(x^{\prime})
\pi_{n-1}(x)}{h_{n-1}(x-x^{\prime})}\ .
\end{equation}
The basis of our analysis relies on the Fokas-Its-Kitaev formulation \cite{FIK0,FIK1} of OPs in terms of the following RHP for the $2\times 2$ matrix $Y(z)$
 (for brevity, we drop the explicit dependence of $Y$ on $n$)
\bea
Y_+(x) = Y_{-}(x) \le[
\begin{array}{cc}
1 & {\rm e}^{-\frac NTV(x)}\cr
0&1
\end{array}
\ri]\ ,\qquad
Y(z) \sim\big(\1 + \mathcal O(z^{-1})\big) \le[\begin{array}{cc}
z^n &0\cr 0&z^{-n}
\end{array}\ri],
\label{OPRHP2}
\eea
and the polynomial  $\pi_n(z) $ is simply $Y_{11}(z)$, while the kernel is recovered from
\be
K_{n,N}(x,x') = e^{-\frac{1}{2}\frac{N}{T}(V(x)+V(x^{\prime}))}\frac{ \le[Y^{-1}(x) Y(x')\ri]_{21}}{-2\pi{\rm i}(x-x')}\ .
\ee
Then the $m$-point joint probability distribution function can be written as the
determinant of the kernel (\ref{eq:kernel}) \cite{Dyson}, \cite{MehtaBook},
\cite{Porter}
\begin{equation*}
\mathcal{R}_{m}^{(n,N)}(x_1,\ldots,x_m):=\det\left(K_{n,N}(x_j,x_k)\right)_{1\leq
j,k\leq m}
\end{equation*}

In the limit $\lim_{n,N\rightarrow\infty}\frac{n}{N}=1$, the
eigenvalue density $\frac{\mathcal{R}^{(n,N)}_1(x)}{n}$ of the
ensemble (\ref{eq:rm}) is asymptotic to the \it equilibrium measure
\rm $\rho(x)$ \cite{Deift}, \cite{Johansson}, \cite{SaffTotik}:
\begin{equation*}
\lim_{n,N\rightarrow\infty,\frac{N}{n}\rightarrow 1}\frac{\mathcal{R}_1^{(n,N)}(x)}{n}=\rho(x),
\end{equation*}
where the $\rho(x)dx=d\mu_{min}(x)$ is the normalized density of the unique
measure $\mu_{min}(x)$ that minimizes the energy
\begin{equation*}
I(\mu)=-T\int_{\mathbb{R}}\int_{\mathbb{R}}\log|x-y|d\mu(x)d\mu(y)+\int_{\mathbb{R}}V(x)d\mu(x)
\end{equation*}
among all Borel probability measures $\mu$ on $\mathbb{R}$. The fact
that $\mu_{min}(x)$ admits a probability density follows from the
assumption that $V(x)$ is real and analytic \cite{McLaughlinDeiftKriecherbauer}. Moreover, it
was shown ibidem that for real and analytic $V(x)$, the
equilibrium measure is supported on a finite union of intervals.

\subsection{Colonization at an outpost}
The following conditions are satisfied by the equilibrium density
$\rho(x)$ \cite{Deift}, \cite{SaffTotik}
\begin{equation}\label{eq:ineq}
\begin{split}
&2T\int_{\mathbb{R}}\log |x-s|\rho(s)ds-V(x)=\ell , \quad
x\in\textrm{Supp}(\rho),\\
&2T\int_{\mathbb{R}}\log|x-s|\rho(s)ds-V(x)\leq \ell, \quad
x\in\mathbb{R}/\textrm{Supp}(\rho),
\end{split}
\end{equation}
for some constant $\ell$ (also known as {\em Robin's constant}). 
For a generic potential $V(x)$, the
inequality in (\ref{eq:ineq}) is satisfied strictly outside the support.  Suppose however that there is some point $x_0\notin\textrm{Supp}(\rho(x))$ where the inequality is not strict
\begin{equation}\label{eq:vcrit}
\begin{split}
&2T\int_{\mathbb{R}}\log|x_0-s|\rho(s)ds-V(x_0)= \ell.
\end{split}
\end{equation}
Such a potential $V$ is called {\bf irregular} (\cite{DKMVZ}); 
a small perturbation of  the potential may induce  a new interval of support of $\rho$ to form around $x_0$. We may think of this phenomenon
as the eigenvalues colonizing the point $x_0$, which we will call 
the {\bf outpost}. This situation has been considered previously and the
term `birth of new cut' was used in some of the studies \cite{Claeys}, \cite{EynardBirth}, \cite{MoBirth}.

In the studies \cite{BertoLee1, BertoLee2}, ,\cite{Claeys}, \cite{EynardBirth}, \cite{MoBirth}, the
colonization phenomenon was considered when a finite number of
eigenvalues start appearing in the outpost $x_0$. It was shown that
the eigenvalue statistics near the outpost can be described by that
of a finite size Hermitian matrix ensemble, or  a {\em microscopic} ensemble.

\subsection{Genus transition in random matrix models: the proliferation of a colony}

\begin{wrapfigure}{r}{0.4\textwidth}
\vspace{0cm}
\resizebox{0.4 \textwidth}{!}{
\input{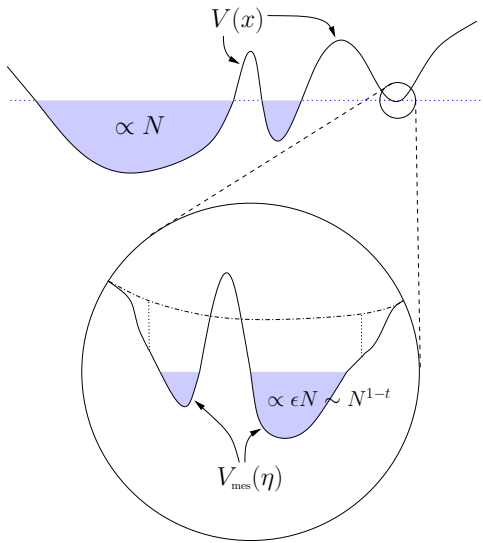}}
\caption{ Mesoscopic colonization.}\label{view}
\vspace{-0.5cm}
\end{wrapfigure}

Some
interesting questions about how new intervals in the support are
forming remained unanswered. For example, whether several intervals in
the support can form simultaneously or they have to form one after
another; how to describe the eigenvalue statistics when the number
of eigenvalues in the colony becomes large. This paper aims at
addressing some of these questions, namely, we want to analyze the transition
from the `first colonization' to the situation
where one or more  new intervals are fully formed. A schematic view is shown in Fig. \ref{view}.  Since the size of the colony, though small compared to the main cut, is taken $o(N)$ and unbounded,  we use the term {\bf mesoscopic colonization}.

The setup is as follows: let $V(x)$ be a critical potential such
that (\ref{eq:vcrit}) is satisfied at a point  $x_0$ outside of the
support of the equilibrium measure and let the order of vanishing of
(\ref{eq:vcrit}) be $2\nu+2$. In particular, at the outpost, we have
\begin{equation}\label{eq:phi}
\phi(x)=V(x)-2T\int_{\mathbb{R}}\log|x-s|\rho(s)ds+l=C_0(x-x_0)^{2\nu+2}(1+\mathcal{O}((x-x_0))\
,\ \ C_0>0\ .
\end{equation}
Without loss of generality we will perform a translation of the problem so that $x_0=0$, but we will keep referring to it at $x_0$ not to confuse it with other zeroes. The function $\phi(x)$ is called {\bf effective potential} since it represents the sum of the external potential $V$ and the Coulomb (two-dimensional) potential generated by the equilibrium distribution.
Let ${\bf B}_J(x)$ be a bump function that is 1 inside an interval
$J\subset\mathbb{R}$ around $x_0$ and $0$ outside an interval
$\tilde{J}\supset J$ around $x_0$. Both $J,\wt J$ are chosen small enough so as not to contain any point of the support of $\rho$. We will study perturbed model 
\begin{equation}\label{eq:prob}
Z_{n,N}^{-1}\exp\left(-\frac{N}{T}\tr \widehat{V}(M)\right)dM,\quad
Z_{n,N}=\int_{\mathcal{H}_n}\exp\le(-\frac{N}{T}\tr \widehat{V}(M)\ri)dM,
\end{equation}
where $\wh {V}(x)$ is a 1-parameter perturbation of $V(x)$ (See
Figure \ref{view})
\begin{equation}
\wh{V}(x)=V(x)+{\bf B}_J(x)A_{\kappa,N}(x),\label{pertpot}
\end{equation}
where $\kappa$ is of order ${\cal O}(N^{1-t})$ with $0<t<1$, and $A(x)$ is analytic near $x=x_0$.   
Due to
(\ref{eq:phi}), we can define a local parameter $\eta$ inside a finite neighborhood ${\mathbb D}$ around $x_0$ as
follows
\begin{align}\label{eq:eta}
&\eta=\left(\frac{N}{\kappa T}\phi(x)\right)^{\gamma}=
\frac{x}{\epsilon}\big(1+{\cal O}(x)\big),\qquad(x\in{\mathbb D})\ ,
\\&\qquad\epsilon:=\left(\frac{\kappa T}{C_0 N}\right)^{\gamma}\ ,\qquad
\gamma:=\frac{1}{2\nu+2}.
\end{align}

We will show that for a suitable choice of the perturbation function $A(x)$ the eigenvalues of the matrix
model (\ref{eq:prob}) are distributed on micro-cuts in $\D$ whose image on the $\eta$--plane is a collection of at most $\nu+1$ segments. In terms of the coordinate $x$, this support shrinks at  a rate $\mathcal O ( (\kappa /N)^\gamma) =\mathcal O(  N^{-t\gamma})$.

The bump function ${\bf B}_J$ is used to keep the technicalities
to its minimum and is not essential to the construction: changing between two such bump functions will introduce a difference in the description which is exponentially small (as $N\to \infty$) and hence beyond all orders of perturbation. Such a manipulation of the potential is very useful in handling the otherwise complicated ``double scaling limit". The result of this paper can be encompassed in the following theorem:

\begin{theorem}\label{thm:main1}
Let $V$ be real analytic and irregular, with an effective potential vanishing at $x_0$ as dictated in (\ref{eq:phi}) and let $\kappa=\kappa_N$ be a sequence of integers\footnote{The requirement is purely technical and could be disposed of, at the price of complicating the analysis, without changing the result.} such that $\kappa = \mathcal O (N^{1-t})$ for some $1>t>0$. Let $\eta$ be the scaling coordinate near $x_0$  given by eq. (\ref{eq:eta}) and  $V_{\mbox{\scriptsize mes}}(\eta)=\eta^{2\nu+2}+\sum_{j=1}^{2\nu+1}t_j\eta^j$,
$t_j\in\mathbb{R}$ be a real monic polynomial potential of degree $2\nu+2$. Let $\mu_{mes}(\eta)$ be its equilibrium measure minimizing
\begin{equation}
\int_{\mathbb R}{V_{\mbox{\scriptsize mes}}}(\eta)d\mu_{mes}(\eta)-\iint_{{\mathbb R}\times{\mathbb R}}\log|\eta-\eta'|d\mu_{mes}(\eta)d\mu_{mes}(\eta')\ .
\end{equation}
 Let $\D$ be a small
neighborhood  of $x_0$ in the complex plane  on which $\eta$
\eqref{eq:eta} is conformal. Then we can define a  function $A(x) = A_{N,\kappa}(x)$ as in Def. \ref{defA}  to be used in (\ref{pertpot}) that is analytic and bounded on $J$, of order $\mathcal O(\epsilon) = \mathcal O((\kappa/N)^{\gamma}) =\mathcal O(N^{-\gamma t}) $ (and uniformly so w.r.t. $N,\kappa$ and $x\in J$) such
that in the limit $N\rightarrow\infty$, $\kappa=\mathcal O(N^{1-t})$, $0<t<1$,
the eigenvalue density of the matrix model (\ref{eq:prob}) in $\D$ is
asymptotically given by the equilibrium measure $\mu_{mes}(\eta)$ in the
$\eta$-plane.
\end{theorem}
The proof takes up the whole paper.
The explicit form of $A(x)$ is given later in \eqref{Az}.

We will argue in Sec. \ref{singularity} that the -admittedly contrived- construction of the perturbation is in fact sufficient to capture the universal behavior.

The eigenvalues of (\ref{eq:prob}), except the ones on the macroscopic cuts, are on the support of
$d\mu_{mes}(\eta)$. From (\ref{eq:eta}), we see that the lengths of these cuts are of order
$\left(\frac{\kappa}{N}\right)^{\gamma}$.

\begin{remark}Theorem \ref{thm:main1} means that for a suitably chosen perturbation of the
critical potential, the number of micro-cuts that are formed depends
on the mesoscopic  potential $V_{\mbox{\scriptsize mes}}(x)$. In particular, multiple cuts can be
formed simultaneously if the equilibrium measure of $V_{\mbox{\scriptsize mes}}(x)$ is
supported on multiple cuts.
\end{remark}

\section{Equilibrium measure in the  mesoscopic problem}
\label{mesog}

Consider the mesoscopic potential in Theorem \ref{thm:main1} with $F(\eta):=\sum_{j=1}^{2\nu+1}t_j\eta^j$, 
\be
V_{\mbox{\scriptsize mes}}(\eta) = \eta^{2\nu +2} + F(\eta)\ ,\ \ \deg
F(\eta)\leq 2\nu +1\ . \label{mesopotential} \ee
We define as usual the corresponding $g$-function as the logarithmic transform of the equilibrium measure: 
\be g_{\mbox{\scriptsize mes}} (\xi) = \int_\R \ln (\xi-\eta)
\rho_{\mbox{\scriptsize mes}}(\eta)\d \eta \label{eq:mesog}\ee where $\rho_{\mbox{\scriptsize mes}}(\eta)$
is the probability measure on $\R$ that minimizes  the familiar
energy functional \be \mathcal F_{mes}[\rho]:= \int V_{\mbox{\scriptsize mes}}(\eta) \rho_{\mbox{\scriptsize mes}}(\eta) \d \eta + \iint  \ln \frac 1{|\eta -\xi|}
\rho_{\mbox{\scriptsize mes}}(\eta)\rho_{\mbox{\scriptsize mes}}(\xi)\d \eta\d \xi \ee

The support of $\rho_{\mbox{\scriptsize mes}}$ is a finite union of intervals
\cite{McLaughlinDeiftKriecherbauer} and it is possible to see
that in fact it can have at most $\nu+1$ disjoint intervals. The
$g$--function has an expansion for large argument of the form
\be
g_{\mbox{\scriptsize mes}}(\xi) =\ln \xi- \sum_{j=1}^\infty \frac
{b_j}{\xi^j}\ ,\ \ b_j:= (-1)^{j} \int_\R
\rho_{\mbox{\scriptsize mes}}(\eta)\eta^j \d \eta. \ee The mesoscopic equilibrium
measure satisfies the same inequalities as in (\ref{eq:ineq}) with
an appropriate (mesoscopic Robin's) constant $\ell_{\mbox{\scriptsize mes}}$: \bea
V_{\mbox{\scriptsize mes}}(\eta) - 2\Re g_{\mbox{\scriptsize mes}}(\eta) +\ell_{\mbox{\scriptsize mes}} =0\ ,\ \ \eta \in {\rm Supp}(\rho_{\mbox{\scriptsize mes}})\ ,\\
V_{\mbox{\scriptsize mes}}(\eta) -2\Re g_{\mbox{\scriptsize mes}}(\eta) +\ell_{\mbox{\scriptsize mes}} \geq 0\ ,\
\ \eta \not\in {\rm Supp}(\rho_{\mbox{\scriptsize mes}})\ . \eea
We will need the
following truncation of the expansion \bd \label{gtrunc} The {\bf
truncated mesoscopic} $g$--function is defined as 
\be \wh g_{\mbox{\scriptsize mes}}(\eta) = \ln \eta - \sum_{j=1}^{k} \frac {b_{j}}{\eta^j}
=: \ln \eta - f_{\mbox{\scriptsize mes}}(\eta)\ .\label{truncg} \ee
Note that we have defined both $\wh g_{mes}$ and the function $f_{mes}$.
 \ed 
\noindent The minimal level of truncation $k$ will be determined in (\ref{kmin}), but for the time being it is a parameter of our problem.

\section {Singularly perturbed variational problem}

In order to construct the deformation of the original problem so that
we obtain the desired double--scaling limit we need to work a bit more
compared to  \cite{BertoLee1, BertoLee2}. In particular the
global $g$-function will  be modified to a certain extent  because mesoscopic colony is ``big" enough to affect the minimization problem for the macroscopic spectrum.
 
Let $V(x)$ be a real--analytic potential. It is known from  \cite{McLaughlinDeiftKriecherbauer} that the support of the corresponding equilibrium measure consists of a finite union of disjoint finite intervals $\bigsqcup [\a_{2j-1} , \a_{2j}]$.
We define the complexified effective potential by  the formula
\be
\varphi(z):= V(z) - 2T \int_\R \rho(t)\ln (z-t)\d t + \ell\label{complexphi}
\ee
Due to the multivaluedness of the logarithm $\varphi$ is only defined on a simply connected domain, which customarily is chosen as $\C\setminus (-\infty, \max\supp(\rho)]$ \cite{DKMVZ}. If the point $x_0$ belongs to a finite spectral gap $(a,b)$ (a finite connected component of $\R\setminus \supp(\rho)$)
we can alternatively define $\varphi$ as a holomorphic function on $\C \setminus \{ (-\infty,a]\cup[b,\infty)\}$; the only effect in this re-definition is to modify the so--called {\em model problem} (or {\em outer parametrix}) by a constant (in $z$) multiplier. It is more convenient  for the discussion (but not at all crucial) to use a definition of $\varphi$ which is analytic at $x_0$ and  so we will assume this is the case. For example, if $x_0$ lies to the right of $\supp \rho$ then no additional complication arises.

The condition that $V(x)$ is irregular (\ref{eq:phi}) at $x_0\not\in \supp(\rho)$ is translated in terms of $\varphi$ as the condition  $\varphi(x) = C_0\,x^{2\nu+2} (1 + \mathcal O(x))$, $C_0>0$.

We will assume, for simplicity, that all other turning points are {\em simple}, namely at the endpoints of the intervals of the support of $\rho$ we have $\varphi'(x) \sim \wt C (x-\alpha_j)^{\frac 1 2 } (1 + \mathcal O(x-\a_j))$.

The goal of this section is to define a small perturbation to the unperturbed $g$--function (or the unperturbed effective potential) which will serve to normalize --eventually-- the RHP for the corresponding orthogonal polynomials.

\bd
\label{def:scalingparameter}
The mesoscopic conformal scaling parameter $\eta$ is defined by the following equations
\bea
\kappa\, \eta^{2\nu+2} = \frac{N}{T} \varphi(x) \ \Leftrightarrow\
\eta:=\eta(x) = \le(\frac N{\kappa T} \varphi(x)\ri )^\gamma =  \frac x{\epsilon} ( 1 + \mathcal O(x)),\label{scalingparameter}\\
 \gamma := \frac 1{2\nu +2}\ \ ,\ \
\epsilon := \le( \frac{\kappa T} {C_0 N}\ri)^{\gamma}\ .\label{epsilon}
\eea
\ed
The choice of symbol ($\eta$) is made on purpose to match the use of coordinate that was made in the previous section (Sec. \ref{mesog}).
Define the following Laurent polynomial in $x$
\be
f(x/\epsilon) := -\res{z=0} \frac {f_{\mbox{\scriptsize mes}}(\eta(z)) }{ z-x} \d z = \sum_{j=1}^{k} \frac {\beta_j}{ (x/\epsilon)^j} \ ,\ \ f_{\mbox{\scriptsize mes}}(\eta) \hbox{ as in (\ref{truncg})}\label{fe}
\ee
We note that $\beta_j = b_j + \mathcal O(\epsilon)$ are analytic functions near  $\epsilon=0$.

The singularly perturbed minimization problem consists now in  minimizing the following functional
\bea
\mathcal F_\epsilon &\& := \int_{\R} \overbrace{\le(V(t)-2\frac {\kappa T }{N} H_\epsilon(x) \ri)}^{=:\wt V(x)}
 \d \mu(t) +T\int\int \d \mu(t) \d\mu(s) \ln \frac 1{|s-t|} \nonumber \\[10pt]
&\& H_\epsilon(x) := \ln |x/\epsilon| - f(x/\epsilon)\label{HE}\ ,\nonumber \\[10pt]
&\&\int \d \mu(t) = 1- \frac {\kappa}N\ ,\ \ \ {\rm supp}(\mu) \subset \R\setminus J\ .
\label{singpert}
\eea
 where the minimization is taken over the set of Borel measure that is
supported on $\mathbb{R}\setminus J$.
Note that, with the above definition of $H_\epsilon(x)$,  the following property is verified.
\begin{equation}\label{intend}
H_\epsilon(x)-\wh g_{\mbox{\scriptsize mes}}(\eta)={\cal O}(\epsilon)+{\cal O}(x)(1+{\cal O}(\epsilon))\ .
\end{equation}

We point out  that the potentials $\wt V = \wt V(x, \epsilon)$ are admissible on $\R\setminus J$ in the sense of  potential theory (\cite{SaffTotik}) for sufficiently small $\epsilon$: let $\rho_\epsilon$ be the corresponding equilibrium measures. Then we can define the modified $g$-function by
\begin{equation}
\wt g(x):=\int_{\mathbb R}\log(x-t)\rho_\epsilon(t)dt\ ,
\end{equation}
and the modified effective potential by
\begin{equation}
\wt\varphi(x):=\wt V(x)-2T\,\wt g(x)+\wt\ell+T\frac{\kappa}{N}\ell_{\mbox{\scriptsize mes}}\ 
\label{modpot}
\end{equation}
where we have written the Robin constant for the modified minimization problem as $\wt\ell+T\frac{\kappa}{N}\ell_{\mbox{\scriptsize mes}}$ for convenience,  and --by definition-- it is such that its real part of $\wt \varphi$ is zero on ${\rm supp}(\rho_\epsilon)$. Note that for $\kappa=0$  ($\epsilon=0$) , the solution of the variational problem (\ref{singpert}) and the original one over the whole real axis coincide since both fulfill eqs. (\ref{eq:ineq}).\\
We can then apply the results of \cite{KuijlaarsMcLaughlinDensity} to conclude that $\wt V$ is a regular potential (for $\epsilon$ small) on $\R\setminus J$.
In particular we quote the relevant 
\bt[\cite{KuijlaarsMcLaughlinDensity}, Theorem 1.2]
\label{KMthm}
Suppose $V$ and $V_n$, $n=1,2,\dots $ are real analytic external fields on $\R$ such that the following hold:
\begin{enumerate}
\item $V_n$ and the first three derivatives converge to $V$ uniformly on compact subsets of $\R$;
\item The growth condition $\lim_{|x|\to \infty} V_n(x)/\ln|x| = +\infty$ holds uniformly in $n$.
\end{enumerate}
Then the supports of the corresponding equilibrium measures  are uniformly bounded. Furthermore, if $V$ is regular then so are all the $V_n$ eventually.
\et
There are two remarks due at this point 
\begin{itemize}
\item Clearly we can adapt the above theorem to any family of potentials $V_\epsilon$ with the obvious modifications of the statement; in this case an isotopy argument implies that if $V_0$ is regular and has $K$ component in the support of the corresponding equilibrium measure, so happens for  $V_\epsilon$, as long as $\epsilon$ is sufficiently small.
\item Thm. \ref{KMthm} is stated on $\R$ but, reading the proof in \cite{KuijlaarsMcLaughlinDensity}, it  appears that there is no difficulty in replacing $\R$ with  $\R \setminus J$, or even any union of intervals, for what matters. 
\end{itemize}
 Thm. \ref{KMthm} with the above trivial extensions implies that the number of components of the supports for $\rho_\epsilon$ is finite and constant (for $\epsilon$ sufficiently small) and the endpoints are smooth functions of $\epsilon$. In fact it is possible to  derive (nonlinear) differential equations for the endpoints as functions of $\epsilon$. In the appendix we give the result without proof, since it is not necessary to the considerations to follow.

\subsection{Modified orthogonal polynomials}
We choose a small interval $J$ around the outpost that does not contain any other  endpoint.
We will consider the following modified orthogonality relations
\bea\label{ptildeV}
h_n\delta_{nm}=\int_{\R}p_n(x) p_m(x) {\rm e}^{-\frac{N}{T}\wh V(x)} \d x
\eea
where the perturbed potential $\wh V(x)$ was given in \eqref{pertpot}.

For simplicity we will also assume that  $\kappa = \kappa_N$ depends on $N$ in such a way that
\begin{itemize}
\item $\kappa = \kappa_N$ is an {\bf integer};
\item $\kappa_N = \mathcal O(N^{1-t})$, $1>t>0$.
\end{itemize}
Were we to allow $\kappa$ to be non-integer, we would have to complicate the analysis by taking into account that when $\kappa$ crosses the half--integers an improved local parametrix needs to be used as in \cite{BertoLee1}. This would only lengthen (considerably) the paper while providing no further insight into the phenomenon we want to describe.

\subsection {Dressing the RHP with the singularly perturbed $g$--function} 

For the orthogonal polynomials at \eqref{ptildeV} we take the RHP \eqref{OPRHP2} for $Y$ {\em with $\widehat V$ instead of $V$}.

We define
\be
\Psi(z) := {\rm e}^{-\frac {N \wt \ell}{2 T} \sigma_3}
{\rm e}^{-\frac {\kappa \ell_{\mbox{\tiny mes}}}{2} \sigma_3} 
\epsilon^{-\kappa\sigma_3}\, Y (z)\, {\rm e}^{-\kappa \left(H_\epsilon(z)-\frac{\ell_{\mbox{\tiny mes}}}{2}\right) \sigma_3}{\rm e}^{-N \left(\wt g(z)-\frac{\wt\ell}{2T}\right) \sigma_3} \label{Psiproblem}\ .
\ee
The various prefactors of $Y(z)$ above are only to ensure that $\Psi(z) = \1 + \mathcal O(z^{-1})$, the $\epsilon^{-\kappa\sigma_3}$ term coming to compensate the term $\ln (x/\epsilon)$ that appears in $H_\epsilon$ (\ref{HE}).
In this way, the $g$--functions is ``stripping off'' the outer parametrix  from ``all" the zeros including the ones at the outpost.  This approach is different from the one in \cite{BertoLee1,BertoLee2} and actually closer to \cite{MoBirth,Claeys}.  
As a result the jumps on $J$ for $\Psi(z)$ become 
\bea
\Psi_+(x) &=& \Psi_-(x)\le[ \begin{array}{cc}
1 &  {\rm e}^{-\frac{N}{T} \le( \wh V - 2T \,\wt g+ \wt \ell\ri) +\kappa (2H_\epsilon(x)-\ell_{\mbox{\tiny mes}}) }\\
0& 1
\end{array}\ri]
\eea
By virtue of the variational problem that $\wt g$ solves and since $\wh V \equiv V$ outside of $J$, the analysis on the support of $\wt \rho$ can be carried out in {\em verbatim} as in \cite{DKMVZ}, keeping in mind that the endpoints are slowly varying functions of $\kappa/N$. We want to focus on the problem near the outpost, as it contains the whole essence of the  new phenomenon.

In order to have locally the (simplest form of the) RHP for the
mesoscopic potential we need to have 
\begin{align}
\frac NT \le(\wh V(x) -2T\,\wt g(x)+ \wt \ell \ri) -2\kappa H_\epsilon(x) +\kappa \ell_{\mbox{\scriptsize mes}}
&\stackrel{\mbox{\scriptsize set}}{=} \kappa \le(V_{\mbox{\scriptsize mes}}(\eta) - 2 \wh g_{\mbox{\scriptsize mes}}(\eta)+\ell_{\mbox{\scriptsize mes}}\ri)
\\&=  \underbrace{\frac
NT\le( V(x) -2T g(x) + \ell \ri)}_{=\kappa \eta^{2\nu+2}} + \kappa
F(\eta) -2\kappa\wh g_{\mbox{\scriptsize mes}}(\eta)+\kappa\,\ell_{\mbox{\scriptsize mes}}\ .
 \end{align}
Simplifying the above expression, we have
\bd
\label{defA}
The perturbation function $A(z) := A_{\kappa,N}(z)$ to be used in the perturbed potential (\ref{pertpot}) is defined as 
\be 
A(z)=\wh V-V =  - \le(2
T(g-\wt g)+ (\ell-\wt \ell)\ri) +\frac{ \kappa T}N \Big(
F(\eta) + 2  \le( H_\epsilon(x)-\wh g_{\mbox{\scriptsize mes}}(\eta) \ri)\Big) \label{Az}
\ee
\ed
\br
We recall what are the input data in Def. \ref{defA}, so as to make clear the definition is not circular: we need
\begin{itemize}
\item the unperturbed nonregular potential $V(z)$ (with the property (\ref{eq:phi}));
\item the mesoscopic potential  $V_{mes}(\eta) = \eta^{2\nu+2} +F(\eta)$, with $F(\eta)$ an arbitrarily chosen polynomial of degree $2\nu+1$;
\item the order of truncation $k$.
\end{itemize}
The other functions appearing in (\ref{Az}) are $H_\epsilon$ (defined in (\ref{HE})), the truncated $g$-function (Def. \ref{gtrunc})
\er

 In
order to manifest the analytic properties of $A(x) = \wh V -V$ we
point out that the singularities of $H_\epsilon(z)$ and $\wh g_{\mbox{\scriptsize mes}}$
cancel out precisely by \eqref{intend} to give a locally analytic function in the
neighborhood of the outpost. Also the largest deviation is given by the term $\frac{\kappa T}{N}F(\eta)=\epsilon^{2\nu+2} \mathcal O(\epsilon^{-2\nu-1})  = {\cal O}(\epsilon)$.  The deviation $2
T(g-\wt g)+ (\ell-\wt \ell)$ is bounded by ${\cal O}(\epsilon^{2\nu+2})$ in general (see, for instance, the lemma \ref{lemma1}).

With this position for $A$ we have the following RHP in the neighborhood of the outpost
\bea
\Psi_+ &\& = \Psi_-\le[ \begin{array}{cc}
1 & {\rm e}^{-\kappa (V_{\mbox{\scriptsize mes}}(\eta) - 2\wh g_{\mbox{\scriptsize mes}}(\eta)+\ell_{\mbox{\scriptsize mes}})}\\
0&1
\end{array}\ri]\ ,\cr
\Psi(z) &\&= \mathcal O(z^0) {\rm e}^{-\kappa H_\epsilon(z)\sigma_3}=
\mathcal O(\eta^0) {\rm e}^{-\kappa \wh g_{\mbox{\scriptsize mes}}(\eta) \sigma_3}\ ,\
\ \eta\to 0\ .
\label{outpostlocal2}
 \eea
The growth behavior at the origin is obtained from the definition of $\Psi$ \eqref{Psiproblem}.
The reason why the two behaviors at $z=0$ on the second line of (\ref{outpostlocal2}) are equivalent is due to the fact that $H_\epsilon(z)$is precisely the singular part of $\wh g_{\mbox{\scriptsize mes}}$, as follows from (\ref{HE}) and (\ref{fe}).

\subsection{Local parametrix at the mesoscopic colony}
\label{Vmeso}
Let $P_j(\eta) = P_j(\eta;\kappa)$ be the {\em monic} orthogonal polynomials for the (varying) measure ${\rm e}^{-\kappa V_{\mbox{\scriptsize mes}}(\eta)}\d \eta$.
We want to construct an exact solution $R_\kappa(\eta)$ to the jump condition (\ref{outpostlocal2})  such that on $\pa \D$ it behaves as $\1 + \mathcal O(N^{-\alpha})$ for some positive $\alpha$ and uniformly in $\eta$.

Consider the matrix
\be
R_\kappa(\eta) = {\rm e}^{-\frac12\kappa\ell_{\mbox{\tiny mes}}\sigma_3}\le[
\begin{array}{cc}
P_\kappa(\eta)  & \mathcal C[P_\kappa](\eta) \\
\frac {-2i\pi}{h_{\kappa-1}} P_{\kappa-1}(\eta) & \frac {-2i\pi}{h_{\kappa-1}}
\mathcal C[P_{\kappa-1}] (\eta)
\end{array}
\ri]{\rm e}^{-\kappa \wh g_{\mbox{\scriptsize mes}}\sigma_3}{\rm e}^{\frac12\kappa\ell_{\mbox{\tiny mes}}\sigma_3}
\ee
It is immediate to verify that it solves the following jump condition and asymptotic behavior
\bea
&&R_{\kappa}(\eta)_+ = R_\kappa(\eta)_-\le[ \begin{array}{cc}
1 & {\rm e}^{-\kappa (V_{\mbox{\scriptsize mes}}(\eta) - 2\wh g_{\mbox{\scriptsize mes}}(\eta)+\ell_{\mbox{\tiny mes}})}\\
0&1
\end{array}\ri]\ ,\label{Rjump}
\\
\label{Rerror}
&&R_\kappa (\eta) = \1 + \mathcal O\le(\frac 1  \eta\ri)\ \quad \eta\rightarrow\infty. \label{317}\\
&&R_\kappa (\eta)= \mathcal{O}(1)e^{-\kappa
\hat{g}_{mes}\sigma_3},\quad \eta\rightarrow 0\ ,
 \eea
 The last error term ${\cal O}(\eta^{-1})$ in (\ref{317}) depends on $\kappa$ and
we need to control this uniformly in the limit $\kappa\rightarrow
\infty$. To do this,
which is a crucial fact to the error analysis of the asymptotics, we
use the standard knowledge on the asymptotic behavior
of the mesoscopic orthogonal polynomials, that comes from the steepest descent analysis \cite{DKMVZ} of the local RHP with respect to the local coordinate $\eta$. Briefly this amount to say
that, as $\kappa \to \infty$ we can obtain expressions
 \be
 R_\kappa (\eta)=\left({\1}+{\cal
O}\left(\frac{1}{\kappa^{E_{\mbox{\scriptsize
mes}}}}\right)\right)\Theta(\eta){\rm e}^{\kappa
g_{\mbox{\scriptsize mes}}(\eta)\sigma_3}{\rm e}^{-\kappa \wh
g_{\mbox{\scriptsize mes}}(\eta)\sigma_3},\quad
\eta\rightarrow\infty.\label{336} 
\ee
Note that mesoscopic Robin's constant disappears by virtue of our
well--crafted choice of perturbation. 

The first factor in (\ref{336}) comes from the error
of the mesoscopic error matrix and $E_{\mbox{\scriptsize mes}}$ is
determined by the nature of the mesoscopic system; for a usual
situation with regular mesoscopic potential, we have
$E_{\mbox{\scriptsize mes}}=1$.  The  $2\times 2$ matrix here denoted by  $\Theta(\eta)$ is the
theta function expression for the asymptotic of the orthogonal
polynomials that solves the so--called ``model problem'', with jumps on the support of $\mu_{mes}$ and in the interval between (See \cite{DKMVZ} or \cite{BertolaMo}). It is known that   the factor $\Theta(\eta)$ behaves as
\begin{equation*}
\Theta(\eta)=\1+\mathcal{O}(\eta^{-1}),\quad
\eta\rightarrow\infty.
\end{equation*}
where the error term is also bounded in $\kappa$ as
$\kappa\rightarrow\infty$.

The trailing exponential factors in \eqref{336} determine the minimal order of the truncation (\ref{truncg}): using Definition \ref{gtrunc},
we have $g_{\mbox{\scriptsize mes}}(\eta)-\wh g_{\mbox{\scriptsize mes}}(\eta)={\cal O}(1/\eta^{k+1})$.
Therefore, for $\eta$ on the boundary $\pa \D$ we have the uniform estimate
\begin{align}
R_\kappa(\eta) &= \left({\1}+{\cal O}\left(\frac{1}{\kappa^{E_{\mbox{\scriptsize mes}}}}\right)\right)\left({\1}+{\cal O}\left(\frac{1}{\eta}\right)\right)\le(\1 + \mathcal O\le(\frac \kappa {\eta^{k+1}}\ri)\ri) \cr
&=\left({\1}+{\cal O}\left(\frac{1}{\kappa^{E_{\mbox{\scriptsize mes}}}}\right)\right)\left({\1}+{\cal O} \le(\frac {\kappa^\gamma }{N^\gamma}\right)\right)\le(\1 + \underbrace{\mathcal O\le(\frac{\kappa^{\gamma(k+1)+1}}{N^{\gamma(k+1)}}\ri)}_{\ds\star}\ri)\ .\label{errorterm}
\end{align}
The last contribution to the error term marked with $\star$ in (\ref{errorterm})  is the most important one: demanding that the error decays imposes a condition  on the minimal $k$ of the truncation in Def. (\ref{gtrunc}). Indeed,  the growth of  $\kappa$ must be  such that there exists a minimal $k_{min}$ for which the last term is $o(1)$. In other words the order of growth of the colony must be
\be
\kappa <N^{1 - \frac 1{\gamma(k+1)+1}}.\label{minorder}
\ee
for $k$ sufficiently large.

If $\kappa  = \mathcal O(N^{1-t})$ for some $0<t<1$  then we need to choose $k$ so that 
\be
\frac 1{\gamma(k+1)+1}<t \ \Leftrightarrow\ \ k>(2\nu+2) \le(\frac 1 t -1\ri)- 1
\label{kmin}
\ee

This determines the minimal order of truncation in (\ref{gtrunc}) and in all the analysis that followed.
 The error bound is then dominated by the last term in (\ref{errorterm}) and can be made as close as desired to  $\mathcal O(\kappa^\gamma /N^\gamma)$ by choosing $k$ sufficiently large.
For instance, choosing the next-to-minimal $k$ --which we do henceforth-- yields an error bound $1 + \mathcal O(N^{-\gamma t})$.

\section{Outer and local parametrices near the turning points}
We will not go into much detail regarding the rest of the asymptotic analysis because it is quite standard.

In fact the strong asymptotic for $\Psi$ in (\ref{Psiproblem}) is obtained in the identical way as in \cite{DKMVZ} by ``opening the lenses'' around the intervals of the support for the perturbed variational problem (\ref{singpert}). As we have assumed, the variational problem is regular outside of a the $\delta$-neighborhood of the outpost, hence the procedure is {\em verbatim} as in \cite{DKMVZ}. The only caveat is that the endpoints are slowly varying functions of the small parameter $ \frac \kappa N = \mathcal O(N^{-t})$.

It should be clear that the error term of the analysis becomes $\mathcal O(1)$ as $t\to 0_+$, namely, as $\kappa$ grows at the same order as $N$ (at which point the new gaps must be ``fully formed''): more and more terms need to be added to the truncation \ref{gtrunc}. Eventually one must solve an exact minimization problem when the colony is fully grown and the transition from mesoscopic to macroscopic will be complete.
\begin{wrapfigure}{r}{0.76\textwidth}
\resizebox{0.76\textwidth}{!}{\input{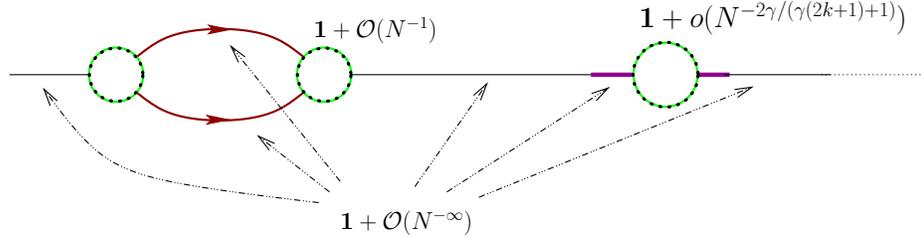}}
\caption{The order of the jumps of  typical residual Riemann--Hilbert problem for the error term, here depicted for a one-cut situation with regular endpoints (where the Airy local parametrix can be used).}
\end{wrapfigure}
The outer parametrix is the same as for the unperturbed problem \cite{DKMVZ}, as well as all the local parametrices near the endpoints of the spectrum.

In particular, denoting by $\Psi$ the outer parametrix, the global parametrix will be of the form
\bea
\Psi^{\mbox{\scriptsize as}} (z)= \le\{
\begin{array}{cc}
\Psi^{\infty} (z) & \text{ outside the local regions}\\
\Psi^{\infty} (z) \mathcal A(z) & \text { near the endpoints}\\
\Psi^{\infty} (z) R_\kappa(z) & \text {at the outpost}
\end{array}
\ri.\ ,
\eea
where $\mathcal A(x)$ is the parametrix constructed by Airy functions as in \cite{DKMVZ}.
Here $\Psi^{\infty}$ is the outer global parametrix constructed out of theta functions as in \cite{DKMVZ} or equivalently using spinors as in \cite{BertolaMo}.

The error matrix $\mathcal E:= \Psi^{\mbox{\scriptsize as}}  \Psi^{-1}$ has several residual jumps: in particular on the boundary of the disk at the outpost we have
\be
\mathcal E_+(\mathcal E_-)^{-1} = \Psi^{\mbox{\scriptsize as}} R_\kappa (\Psi^{\mbox{\scriptsize as}})^{-1}  = \1 + {\cal O}(N^{-\gamma t})\ .
\ee

We now comment on why the choice of bump function is totally irrelevant; indeed, on the real axis and outside of $\D$ the jump is exponentially close to the identity matrix (uniformly). Changing bump function trades such jump by another one, equally close to the identity jump, while leaving the jumps identical within $\D$. As a consequence, the ratio of the solutions corresponding to different choices of bump functions would solve a RHP with jumps exponentially close to the identity everywhere (and in $L^\infty\cap L^2$). Thus the two solutions would differ only by exponentially suppressed terms, well beyond any order of perturbation.

\subsection{Universality of the behavior}
\label{singularity}
The perturbation of the potential has been chosen in the contrived  form (\ref{pertpot}, Def. \ref{defA}) to eventually yield the simplest form for the local Riemann--Hilbert problem in Sec. \ref{Vmeso}; the gist of all the construction is such that in the scaling coordinate $\eta$ the jump on the interval $J$ is  given precisely by (\ref{Rjump}).
As often happens \cite{BertoLee1,BertoLee2}, the logic of our construction is slightly backwards from the more conventional approaches \cite{EynardBirth,MoBirth, Claeys}: we guess what local RHP would give the the phenomenon we expect on heuristic grounds, and try to ``reverse-engineer'' the appropriate deformation of the potential. This approach, while completely rigorous and also simpler to implement, is possibly not the most transparent to the reader.
The perturbation $A(z)$ (\ref{defA}) is {\bf(a)} analytic in $z$ and {\bf (b)} of order $(\kappa/N)^{\frac 1{2\nu+2}}$: from a heuristic point of view (based also on similar setups in the study of the universal unfolding of singularities \cite{Arnoldsing}) it is natural to expect that these are the only relevant features to generate a mesoscopic colonization.  Of course there is much more detailed information that goes into our approach, because the mesoscopic colonization as we described---{\em with a fixed (i.e. non-scaling in $N$) local matrix model}---can only be obtained as a multi-scaling limit; isolating $\kappa$ eigenvalues requires one scaling, forming a specific local cut structure will require rather complicated scalings.   While we could not find a simpler, more direct path that starts from the perturbation and ends at a full description of the scaling regime, we do not expect that such a description, while logically more appealing, would be any simpler.
\section{Conclusion and generalizations}
\begin{itemize}
\item A quite parallel analysis could be performed in the case of the colonization of a hard-edge as in \cite{BertoLee2}. While the logic is identical, there are sufficient small details that would require a separate analysis, but with the final picture being  completely analogous: in that case too one can have --depending on the degree of irregularity of the unperturbed potential-- a mesoscopic growth of several meso-intervals for the equilibrium measure. We believe that the analysis is not sufficiently different to require a separate paper and yet not similar enough to put it here at the expense of clarity and conciseness.

\item It was also  pointed out to one of us\footnote{We thank B. Dubrovin and T. Grava for the indication.}  that the technique of analyzing the colonization  (microscopic and mesoscopic) can be applied almost verbatim to the study of the trailing-edge of the solution of the Korteweg-deVries equation after the time of gradient catastrophe; we reserve to come back on this issue on a subsequent publication.

\item Since the mesoscopic potential can be an arbitrary polynomial, we could choose $V_{\mbox{\small mes}}$ as a nonregular potential such that it has a point outside the support of $\rho_{\mbox{\small mes}}$ where the variational  inequalities (\ref{eq:ineq}) fail. Thus, one may have the whole picture of microscopic/mesoscopic colonization within the analysis of the mesoscopic parametrix. By perturbing $V_{\mbox{\small mes}}$ accordingly one could study a {\bf multiscale colonization}. Since a polynomial potential of degree $2d+2$ can have such a nonregular point with order at most $2d$, we can ``embed'' the micro/mesoscopic pictures one into another at several nested scales at most $2\nu$ times, if the macroscopic potential has an irregular point as the one studied in this paper. We could call this multiscale situation the ``Matryoshka\footnote{A Matryoshka doll is Russian toy consisting in a set of dolls of decreasing sizes placed one inside the other.} colonization''.
\end{itemize}

\appendix
\section{Differential equations for the endpoints}

Here we simply state a result (Prop. \ref{CorODE})  that can be proved along the lines of the  Buyarov-Rakhmanov equation
We use the same notation as in the text (\ref{modpot}) and we set 
\be
y = \frac 1 2 \wt V'(x) - T\int \frac {\rho_\epsilon (s)\d s}{s-x} = \wt \phi'(x).
\ee
It is known that $y$ solves a (pseudo) algebraic equation of the form 
\be
y^2 = F_\epsilon(x)^2 \prod_{j=1}^{2g+2} (x-\alpha_j(\epsilon))
\ee
and $F_\epsilon(x)$ is a real--analytic function (depending on $\epsilon$) with a pole of degree $k$ at $x=0$.
\bl
\label{lemma1}
For small nonnegative  $\epsilon$ we  have (the dot means differentiation in $\epsilon$)
\be
\omega(x) \d x := \dot y\d x
\ee
is the unique meromorphic differential (whose existence follows from standard arguments in algebraic geometry) on the hyperelliptic Riemann surface branched at the endpoints $w^2 = \prod_{j=1}^{2g+2} (x-\alpha_j)$ with the properties that
\begin{enumerate}
\item all periods are purely imaginary;
\item $\omega(x)\d x$ has poles only at the point above $x=0$ with residues $\mp2T$ (respectively, on each sheet).
\item at $x=0$ (on the physical sheet) it behaves as
\be
\omega(x)  \sim -\frac {2T} x  + \frac{\gamma T}{C_0\epsilon^{2\nu+1}}\pa_\epsilon \pa_x ( f(x/\epsilon))
\ee
 Note that the second part contains  poles of order strictly higher than one and hence corresponds to a second--kind differential.
\end{enumerate}
In particular we note that $\omega(x)$ can be written as
\be
\omega(x) = \frac {R(x)}{ \sqrt{\prod_{j=1}^{2g+2} (x-\alpha_j)}}
\ee
with $R(x)$ a rational function of the form
\be
R(x) = P_{k+1}\le(\frac 1 x \ri) + P_{g-1}(x)
\ee
and $P_m(Z)$ denotes some polynomial of degree $m$ of the indeterminate $Z$.  The above three facts completely determine $R(x)$ as a function of $\alpha_j$'s, $\beta_j$'s and $\epsilon$.
\el

\bp
\label{CorODE}
The endpoints solve the following differential equation
\bea
\dot \alpha_j  &\& =\frac{R(\alpha_j)}{F(\a_j)\prod_{k\neq j} (\a_j-\a_k)}\ ,\\
\dot F_\epsilon(x)&\& = \frac{R(x) - F_\epsilon(x) \sum_{j} \dot \alpha_j \prod_{k\neq j} (x-\a_k)}{\prod_{j} (x-\a_j)}\ .
\eea
\ep

\bibliographystyle{plain}
\bibliography{Colonization}

\def\cydot{\leavevmode\raise.4ex\hbox{.}}
\begin{thebibliography}{10}

\bibitem{Arnoldsing}
V.~I. Arnol'd.
\newblock Singularities of smooth mappings.
\newblock {\em Uspehi Mat. Nauk}, 23(1):3--44, 1968.

\bibitem{BertoLee1}
M.~Bertola and S.~Y. Lee.
\newblock First colonization of a spectral outpost in random matrix theory.
\newblock {\em Constr. Approx.}, 2008 (in press).

\bibitem{BertoLee2}
M.~Bertola and S.~Y. Lee.
\newblock First colonization of a hard-edge in random matrix theory.
\newblock {\em Constr. Approx.}, In press, 2009.

\bibitem{BertolaMo}
M.~Bertola and M.~Y. Mo.
\newblock Commuting difference operators, spinor bundles and the asymptotics of
  orthogonal polynomials with respect to varying complex weights.
\newblock {\em Adv. Math.}, 220(1):154--218, 2009.

\bibitem{Claeys}
Tom Claeys.
\newblock The birth of a cut in unitary random matrix ensembles.
\newblock {\em Int Math Res Notices}, 2008(article ID rnm166):40 pages, 2008.

\bibitem{McLaughlinDeiftKriecherbauer}
P.~Deift, T.~Kriecherbauer, and K.~T.-R. McLaughlin.
\newblock New results on the equilibrium measure for logarithmic potentials in
  the presence of an external field.
\newblock {\em J. Approx. Theory}, 95(3):388--475, 1998.

\bibitem{DKMVZ}
P.~Deift, T.~Kriecherbauer, K.~T.-R. McLaughlin, S.~Venakides, and X.~Zhou.
\newblock Uniform asymptotics for polynomials orthogonal with respect to
  varying exponential weights and applications to universality questions in
  random matrix theory.
\newblock {\em Comm. Pure Appl. Math.}, 52(11):1335--1425, 1999.

\bibitem{Deift}
P.~A. Deift.
\newblock {\em Orthogonal polynomials and random matrices: a
  {R}iemann-{H}ilbert approach}, volume~3 of {\em Courant Lecture Notes in
  Mathematics}.
\newblock New York University Courant Institute of Mathematical Sciences, New
  York, 1999.

\bibitem{Dyson}
F.~J. Dyson.
\newblock Correlation between the eigenvalues of a random matrix.
\newblock {\em Comm. Math. Phys.}, 19:235--250, 1970.

\bibitem{EynardBirth}
B.~Eynard.
\newblock Universal distribution of random matrix eigenvalues near the ``birth
  of a cut''; transition.
\newblock {\em Journal of Statistical Mechanics: Theory and Experiment},
  2006(07):P07005, 2006.

\bibitem{FIK0}
A.~R. Its, A.~V. Kitaev, and A.~S. Fokas.
\newblock An isomonodromy approach to the theory of two-dimensional quantum
  gravity.
\newblock {\em Uspekhi Mat. Nauk}, 45(6(276)):135--136, 1990.

\bibitem{FIK1}
A.~R. Its, A.~V. Kitaev, and A.~S. Fokas.
\newblock Matrix models of two-dimensional quantum gravity, and isomonodromic
  solutions of {P}ainlev\'e ``discrete equations''.
\newblock {\em Zap. Nauchn. Sem. Leningrad. Otdel. Mat. Inst. Steklov. (LOMI)},
  187(Differentsialnaya Geom. Gruppy Li i Mekh. 12):3--30, 171, 174, 1991.

\bibitem{Johansson}
K.~Johansson.
\newblock On fluctuations of eigenvalues of random {H}ermitian matrices.
\newblock {\em Duke Math. J.}, 91(1):151--204, 1998.

\bibitem{KuijlaarsMcLaughlinDensity}
A.~B.~J. Kuijlaars and K.~T-R McLaughlin.
\newblock Generic behavior of the density of states in random matrix theory and
  equilibrium problems in the presence of real analytic external fields.
\newblock {\em Comm. Pure Appl. Math.}, 53(6):736--785, 2000.

\bibitem{MehtaBook}
Madan~Lal Mehta.
\newblock {\em Random matrices}, volume 142 of {\em Pure and Applied
  Mathematics (Amsterdam)}.
\newblock Elsevier/Academic Press, Amsterdam, third edition, 2004.

\bibitem{MoBirth}
Man~Yue Mo.
\newblock The {R}iemann-{H}ilbert approach to double scaling limit of random
  matrix eigenvalues near the "birth of a cut" transition.
\newblock {\em Int. Math. Res. Not.}, 2008(ID rnn042), 2008.

\bibitem{Porter}
C.~E. Porter, editor.
\newblock {\em Statistical theories of spectra: {F}luctuations, a collection of
  reprints, original papers, with an introductory review}.
\newblock Academic Press (New York), 1965.

\bibitem{SaffTotik}
Edward~B. Saff and Vilmos Totik.
\newblock {\em Logarithmic potentials with external fields}, volume 316 of {\em
  Grundlehren der Mathematischen Wissenschaften [Fundamental Principles of
  Mathematical Sciences]}.
\newblock Springer-Verlag, Berlin, 1997.
\newblock Appendix B by Thomas Bloom.

\bibitem{Szego}
G.~Szeg\"{o}.
\newblock {\em Orthogonal polynomials}, volume XXIII of {\em American
  Mathematical Society Colloquium Publications}.
\newblock American Mathematical Society, Providence, RI, 1975.

\end{thebibliography}

\end{document}